\documentclass[preprint,showpacs,preprintnumbers,amsmath,amssymb]{revtex4} % newly added
\usepackage{amssymb}
\usepackage{graphicx}% Include figure files
\usepackage{dcolumn}% Align table columns on decimal point
\usepackage{bm}% bold math
\usepackage{booktabs}

\begin{document}

\title{Quantum correlation and classical correlation dynamics in the spin-boson model}
\author{Rong-Chun Ge, Ming Gong, Chuan-Feng Li$\footnote{email: cfli@ustc.edu.cn}$, Jin-Shi Xu, Guang-Can Guo}
\affiliation {Key Laboratory of Quantum Information, University of
Science and Technology of China, CAS, Hefei, 230026, People's
Republic of China}
\date{\today }
%\pacs{68.65.Hb, 73.22.-f, 78.67.Hc }% PACS, the Physics and Astronomy
% 68.65.-k Low-dimensional, mesoscopic, and nanoscale systems:
%          structure and nonelectronic properties
% 68.65.Hb Quantum dots
% 73.22.-f Electronic structure of nanoscale materials: clusters,
%          nanoparticles, nanotubes, and nanocrystals
% 73.21.La Electron states and collective excitations in multilayers, quantum
%          wells, mesoscopic, and nanoscale system: Quantum dots
% 78.67.Hc Optical properties of low-dimensional structures: Quantum dots
% 73.22.-f Electronic structure of nanoscale materials: clusters,
%          nanoparticlehell, nanotubes, and nanocrystals
% 71.15.-m computational methodology use 71.15.-m
%nanoparticles, nanotubes, and nanocrystals

\begin{abstract}
We study the quantum correlation and classical correlation dynamics
in a spin-boson model. For two different forms of spectral density,
we obtain analytical results and show that the evolutions of both
correlations depend closely on the form of the initial state. At the
end of evolution, all correlations initially stored in the spin
system transfer to reservoirs. It is found that for a large family
of initial states, quantum correlation remains equal to the
classical correlation during the course of evolution. In addition,
there is no increase in the correlations during the course of
evolution.

\end{abstract}


\maketitle

It is generally accepted that entanglement plays a crucial role in
quantum information processing. However, it is fragile due to the
inevitable interaction of the system with its environment. There
have been many works on entanglement for all kinds of systems
\cite{Zh01,Sa07,Ts07}. Many extraordinary results have been
obtained, such as entanglement sudden death (ESD) \cite{Z01, Yu04,
Yu06}, and sudden birth \cite{V09}.

However, more and more works show that, even without entanglement,
there are quantum tasks \cite{Me03,Da08,L08} superior to their
classical counterparts. This is due to their nonzero quantum
correlation ($Q$). Recently, some special results have been obtained
for $Q$ and its counterpart classical correlation ($C$), such as
sudden change \cite{Ma09,W09}. Therefore, it is important and
interesting to investigate $Q$ and $C$ in more open systems. $Q$,
introduced by Henderson and Vedral \cite{H01}, is identical to
quantum discord introduced by Olliver and Zurek \cite{H02} in the
case of two qubits \cite{Ha04,Luo08,Ma09}. $Q$ describes all
nonclassical correlations in a state, while entanglement is only a
special part of the correlations.

To obtain $Q$, one must acquire $C$ first. As shown in \cite{ Ve03},
$C(\rho_{s_1s_2}) \equiv \max_{\{\Pi_j\}} [S(\rho_{s_1}) -
S_{\{\Pi_j\}} (\rho_{s_1|s_2})]$, where the maximum is taken over
the set of positive-operator-valued measurements (POVM) $\{\Pi_j\}$
in partition $s_2$. $Q$ is then obtained by subtracting $C$ from the
quantum mutual information. In fact, $Q$ is asymmetric \cite{H02},
it depends on which partition the measurement is carried out in.
Only under the condition that $S(\rho_A)= S(\rho_B)$ (where $A$ and
$B$ are the two partitions, and $S(\rho)$ denotes the Von Neumann
entropy of state $\rho$), does it become symmetric \cite{Ve03}. It
is difficult to obtain $Q$ for a general state, owing to the maximal
operation. There are only a few works \cite{Di08,Ma09,W09} on the
dynamics of correlations, and more work is needed to gain a deep
understanding of quantum and classical correlations.

In this Brief Report, we investigate the dynamics of $Q$ and $C$ in
a spin-boson model, which is conventionally used to study open
system. For two different forms of spectral density, we acquire
analytical results for both $Q$ and $C$ for different partitions. It
is shown that evolutions of $Q$ and $C$ depend closely on the form
of the initial state. For a large family of initial states, $Q$
remains equal to $C$ during the course of evolution. We also find
oscillation of $Q$ and $C$. Furthermore, we find that all $Q$ and
$C$ initially stored in the spin system run into the boson
reservoirs continuously without sudden death, which is in sharp
contrast with the case for entanglement. In addition, the absence of
an increase in correlations is investigated.

The model we are considering is two spins interacting independently
with their own boson reservoirs without interaction between the two
spins \cite{Lo08}. The Hamiltonian reads as
\begin{eqnarray}
\hat{H}&=&\sum^2_{i=1}(\frac\omega2 \hat{\sigma}_i^z  + \sum^N_{k=1}
\omega_k\hat{b}^{\dagger}_{i,k}\hat{b}_{i,k})\nonumber\\
&+& \sum^2_{i=1}\sum^N_{k=1} g_{i,k}(\hat{\sigma}_i^-
\hat{b}^{\dagger}_{i,k} + \hat{\sigma}^+_i\hat{b}_{i,k}),
\end{eqnarray}

(with $N\rightarrow\infty$). Here, $\hat{b}_{ik}$ $(i=1,2)$ is an
annihilation operator of the $k$th model in the $i$th reservoir with
corresponding frequency $\omega_k$. $\hat{\sigma}_i^z$,
$\hat{\sigma}_i^+$ ($\hat{\sigma}_i^-$), and $\omega$ represent the
Pauli operator, raising (lowering) operator, and Zeeman splitting of
the $i$th spin, and $g_{i,k}$ denotes the coupling strength between
the $k$th model in the $i$th reservoir and the corresponding spin.
As is well known, all information on the coupling between the system
and reservoir is included in the spectral density, written as
$J(\omega) = \sum_k |g_k|^2 \delta(\omega - \omega_k)$ (Where we
suppose both spins and reservoirs are identical).

Let us first consider the evolution of the initial state with two
excitations in the spin system,
\begin{equation}
|\Phi_0\rangle =
(\alpha|0\rangle_{s_1}|0\rangle_{s_2}+\beta|1\rangle_{s_1}|1\rangle_{s_2})|0\rangle_{r_1}|0\rangle_{r_2},
\end{equation}
with $|\alpha|^2 + |\beta|^2 = 1$, where the collective state
$|0\rangle_{r_i} = \prod_{k=1}^N|0_k\rangle_{r_i}$ indicates there
is no excitation in reservoir $r_i$. Evolution of the whole system
is given by
\begin{eqnarray}
|\Psi_t\rangle &=& \beta(\xi(t)|1\rangle_{s_1}|0\rangle_{r_1}+
\sum_{k=1}^N\lambda_k(t)|0\rangle_{s_1}|1_k\rangle_{r_1})\nonumber\\
&\times&(\xi(t)|1\rangle_{s_2}|0\rangle_{r_2}+
\sum_{l=1}^N\lambda_l(t)|0\rangle_{s_2}|1_l\rangle_{r_2})\nonumber\\
&+&\alpha|0\rangle_{s_1}|0\rangle_{s_2}|0\rangle_{r_1}|0\rangle_{r_2}.
\end{eqnarray}
After introducing the collective state $|1\rangle_{r_i} =
\frac{1}{\chi}\sum_{k=1}^N\lambda_k(t)|1_k\rangle_{r_i}$ with one
excitation in the reservoir, the reduced density matrix of the spin
system is
\begin{eqnarray}
\rho_{s_1s_2}(t) =
\begin{pmatrix}
|\alpha|^2 + |\beta\chi^2|^2 &0 &0 &\alpha\beta^*\xi^{*2}\\
0 &|\beta\xi\chi|^2 &0 &0\\
0 &0 &|\beta\xi\chi|^2 &0\\
\alpha^*\beta\xi^2 &0 &0 &|\beta\xi^2|^2
\end{pmatrix}.
\end{eqnarray}

In our case, we find that the maximum can be obtained at
$\theta=\pi/4$ \cite{Ma09}; therefore, the analytical results of $C$
and $Q$ are
\begin{eqnarray}
C(\rho_{s_1s_2}(t))
&=&Q(\rho_{s_1s_2}(t))=\text{H}(|\beta\xi|^2)\nonumber\\
&-& \text{H}(\frac12 (1+\sqrt{1-4|\beta\xi\chi|^2})),
\end{eqnarray}
where we have introduced the Shannon entropy $\text{H}(x)
=-x{\textsf{log}}_2x-(1-x){\textsf{log}}_2(1-x)$. For the
reservoirs, $C$ and $Q$ read as
\begin{eqnarray}
C(\rho_{r_1r_2}(t))&=& Q(\rho_{r_1r_2}(t))=\text{H}(|\beta\chi|^2
)\nonumber\\
&-& \text{H}(\frac12 (1+\sqrt{1-4|\beta\xi\chi|^2})).
\end{eqnarray}

First, when the spectral density is flat,
\begin{equation}
J(\omega) = \gamma,
\end{equation}
where $\gamma$ is a constant that is commonly used as the Markov
approximation with the interval of the spectral density much broader
than the corresponding energy scale of the system. We obtain
parameters
\begin{equation}
\chi = \sqrt{1- e^{-\gamma t}}, \quad \quad \xi = e^{-\gamma t/2}.
\end{equation}

The concurrence \cite{Wo98} of the system is $\max\{0,e^{-\gamma t}
(|\alpha\beta|-|\beta|^2(1-e^{-\gamma t}))\}$. If
$|\alpha|<|\beta|$, there will be ESD \cite{Lo08}. As time
approaches infinity, the asymptotic behavior of $C$ for the spin
system $\rho_{s_1s_2}$ becomes
\begin{equation}
\lim_{t\rightarrow\infty}C(\rho_{s_1s_2}(t)) \sim (|\beta|^2 -
|\beta|^4) \gamma t e^{-2\gamma t}/\ln2,
\end{equation}
which does not depend on the initial value $C(\rho_{s_1s_2}(0)) = -
(1 - |\beta|^2) {\textsf{log}}_2 (1- |\beta|^2 )- |\beta|^2
{\textsf{log}}_2 |\beta|^2$. It is shown that $C$ of the spin system
is depleted gradually without sudden death.
\begin{figure}
\includegraphics[width=3in]{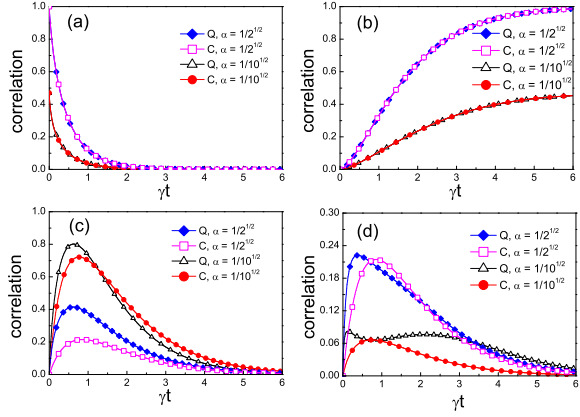}
\caption{(Color online) Evolution of $Q$ and $C$ among different
partitions with flat spectral density for the initial state
$(\alpha|0\rangle_{s_1}|0\rangle_{s_2}
+\beta|1\rangle_{s_1}|1\rangle_{s_2})|0\rangle_{r_1}|0\rangle_{r_2}$.
Blue diamonds and magenta squares denote $Q$ and $C$ for the Bell
state' respectively, while dark triangles and red circles denote $Q$
and $C$ for $\alpha=1/\sqrt{10},\beta=3/\sqrt{10}$. (a) Spin system
$\rho_{s_1s_2}$. (b) Reservoirs $\rho_{r_1r_2}$. (c) Spin $s_1$ with
its corresponding reservoir $r_1$, $\rho_{s_1r_1}$. (d) Spin $s_1$
with the other reservoir $r_2$, $\rho_{s_1r_2}$.}
\end{figure}
\begin{figure}
\includegraphics[width=3in]{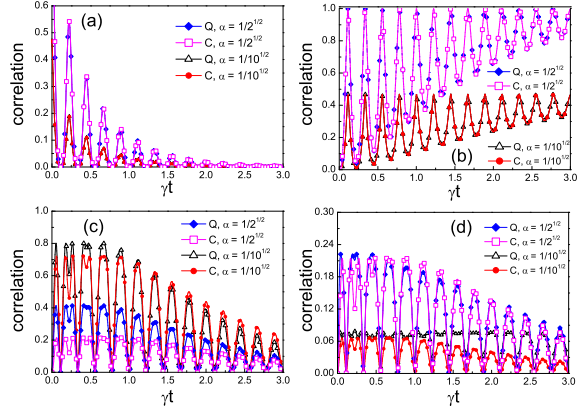}
\caption{(Color online) Dynamics of $Q$ and $C$ for the initial
state $(\alpha|0\rangle_{s_1}|0\rangle_{s_2}
+\beta|1\rangle_{s_1}|1\rangle_{s_2})|0\rangle_{r_1}|0\rangle_{r_2}$
and the spectral density taking the Lorentz form $W/\lambda =
\sqrt{200}$ . Blue diamonds and magenta squares denote $Q$ and $C$
for the Bell state, respectively, while dark triangles and red
circles denote $Q$ and $C$ for
$\alpha=1/\sqrt{10},\beta=3/\sqrt{10}$. (a) Spin system
$\rho_{s_1s_2}$. (b) Reservoirs $\rho_{r_1r_2}$. (c) Spin $s_1$ with
its corresponding reservoir $r_1$, $\rho_{s_1r_1}$. (d) Spin $s_1$
with the other reservoir $r_2$, $\rho_{s_1r_2}$.}
\end{figure}

According to Eq. (5), $C$ is equal to $Q$ during the evolution. As
shown in Figure 1(a), with $\alpha=1/\sqrt{2}, \beta=1/\sqrt{2}$,
and $\alpha =1/\sqrt{10}, \beta= 3/\sqrt{10}$, the correlations tend
to zero monotonically and continuously.

According to Eq. (6) for reservoirs, $C$ equals $Q$, and their
asymptotical behaviors are described by
\begin{equation}
\lim_{t\rightarrow\infty}Q(\rho_{r_1r_2}(t)) = C(\rho_{r_1r_2}(t))
\sim Q(\rho_{s_1s_2}(0)).
\end{equation}

Figures 1(a) and (b) show that both $Q$ and $C$ initially stored in
the spin system run into reservoirs gradually as the whole system
achieves equilibrium. Evolutions of $Q$ and $C$ between the system
and reservoirs may reflect the transference processes of $Q$ and $C$
between the system and reservoirs. $s_1$ ($s_2$) becomes correlated
with $r_1$ ($r_2$) owing to their interaction as shown in Fig. 1(c),
therefore, $r_1$ tangles with $r_2$ through $s_1$ and $s_2$. At the
end of evolution there is no entanglement between reservoirs and
system; therefore, the system can be discarded without any effect,
and all correlations transfer to reservoirs.

We also find that $Q$, $C$, concurrence between $s_1$ amd $r_1$, and
concurrence between $s_2$ and $r_2$, do not increase during the
course of evolution. They are measured by the square of correlations
$Q(t)^2 \equiv
Q(\rho_{s_1s_2}(t))^2+Q(\rho_{s_1r_2}(t))^2+Q(\rho_{s_2r_1}(t))^2+Q(\rho_{r_1r_2}(t))^2\leq
Q(0)^2$, $C(t)^2 \equiv
C(\rho_{s_1s_2}(t))^2+C(\rho_{s_1r_2}(t))^2+C(\rho_{s_2r_1}(t))^2+C(\rho_{r_1r_2}(t))^2\leq
C(0)^2$, and $Con(t)^2\equiv
Con(\rho_{s_1s_2}(t))^2+Con(\rho_{s_1r_2}(t))^2+Con(\rho_{s_2r_1}(t))^2+Con(\rho_{r_1r_2}(t))^2\leq
Con(0)^2$. It can be seen that for partitions $s_1$ and $s_2$,
interaction between $s_2$ and $r_2$ is simply a local unitary
operation; therefore, correlations will not exceed those initially
stored between $s_1$ and $s_2$.

Until now, we have mainly focused on the weak coupling regime, where
dynamics of correlations are monotonic and irreversible. Things will
be different in the strong coupling regime. We consider spectral
density taking the Lorentz form, which is usually encountered in a
cavity \cite{Sa08}. Here we take the spectral density centered at
$\omega_0$ (resonance with the spin system):
\begin{equation}
J(\omega) = (W^2\lambda/\pi)\frac{1}{(\omega-\omega_0)^2 +
\lambda^2},
\end{equation}
where $1/\lambda$ is used to measure the correlation time of the
reservoir. For $\lambda\sim 0$, we obtain $J(\omega)\sim
W^2\delta(\omega-\omega_0)$, which displays a strong non-Markov
effect. We have the parameters
\begin{eqnarray}
\xi &=& e^{-\lambda t/2}[\lambda/\sqrt{\lambda^2 -
4W^2}\sinh(\sqrt{\lambda^2 - 4W^2}t/2)\nonumber\\
&+&\cosh(\sqrt{\lambda^2 -
4W^2}t/2)],\nonumber\\
\chi &=& \sqrt{1-\xi^2}.
\end{eqnarray}

Concurrence  due to the strong non-Markov effect (measured by
$W/\lambda$) is given by
$\max\{0,|\alpha^*\beta\xi^2|-|\beta^2\xi^2\chi^2|\}$, which has
oscillation. $C$ and $Q$ oscillate too, as shown in Fig. 2(a).
According to Eq. (5), for this kind of initial state, although the
coupling has changed, $C$ is equal to $Q$. As shown in Fig. 2(a),
the evolutions of $Q$ and $C$ are continuous. Evolutions of $Q$ and
$C$ for reservoirs also have oscillation, as shown in Fig. 2(b).
From the analytical results, we find that both $Q$ and $C$ initially
stored in the spin system transfer to the reservoirs over time.
Figures 2(c) and 2(d) show the evolutions of correlations for
$\rho_{s_1r_1}$ and $\rho_{s_1r_2}$, respectively. A non increase in
correlations and concurrence is noted, but both correlations and
concurrence oscillate.

\begin{figure}
\includegraphics[width=3in]{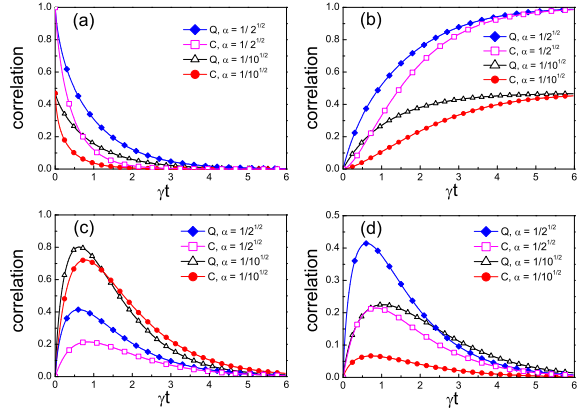}
\caption{(Color online) Evolution of $Q$ and $C$ for the initial
state $(\alpha|0\rangle_{s_1}|1\rangle_{s_2} +
\beta|1\rangle_{s_1}|0\rangle_{s_2})|0\rangle_{r_1}|0\rangle_{r_2}$
with both reservoirs having a flat spectrum. Blue diamonds and
magenta squares denote $Q$ and $C$ for the Bell state, respectively,
while dark triangles and red circles denote $Q$ and $C$ for
$\alpha=1/\sqrt{10},\beta=3/\sqrt{10}$. (a) Spin system
$\rho_{s_1s_2}$. (b) Reservoirs $\rho_{r_1r_2}$. (c) Spin $s_1$ with
reservoir $r_1$, $\rho_{s_1r_1}$. (d) Spin $s_1$ with reservoir
$r_2$, $\rho_{s_1r_2}$.}
\end{figure}
\begin{figure}
\includegraphics[width=3in]{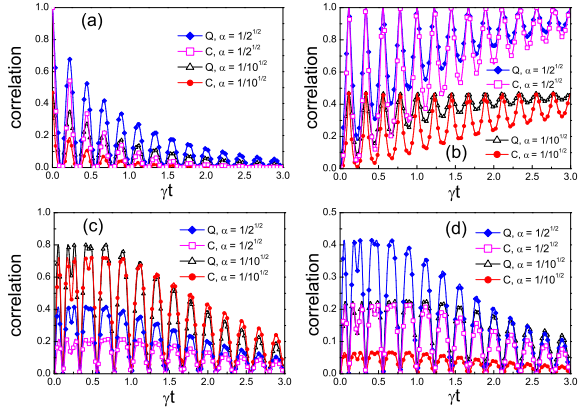}
\caption{(Color online) Dynamics of $Q$ and $C$ for the initial
state $(\alpha|0\rangle_{s_1}|1\rangle_{s_2} +
\beta|1\rangle_{s_1}|0\rangle_{s_2})|0\rangle_{r_1}|0\rangle_{r_2}$
with the spectral density of reservoirs taking the Lorentz form
$W/\lambda = \sqrt{200}$. Blue diamonds and magenta squares denote
$Q$ and $C$ for the Bell state, respectively, while dark triangles
and red circles denote $Q$ and $C$ for
$\alpha=1/\sqrt{10},\beta=3/\sqrt{10}$. (a) Evolution of spin system
$\rho_{s_1s_2}$. (b) Reservoirs $\rho_{r_1r_2}$. (c) Spin $s_1$ with
reservoir $r_1$, $\rho_{s_1r_1}$. (d) Spin $s_1$ with reservoir
$r_2$, $\rho_{s_1r_2}$.}
\end{figure}

We then consider the initial state with only one excitation in the
spin system:
\begin{equation}
|\Phi_0\rangle = (\alpha |0\rangle_{s_1} |1\rangle_{s_2} + \beta
|1\rangle_{s_1} |0\rangle_{s_2}) |0\rangle_{r_1} |0\rangle_{r_2}.
\end{equation}

Evolution of the spin system is given by the reduced density matrix:
\begin{eqnarray}
\rho_{s_1s_2}(t) =
\begin{pmatrix}
|\chi|^2 &0 &0 &0\\
0 &|\alpha\xi|^2 &\alpha\beta^*|\xi|^2 &0\\
0 &\alpha^*\beta|\xi|^2 &|\beta\xi|^2 &0\\
0 &0 &0 &0
\end{pmatrix}.
\end{eqnarray}

Interestingly, in this situation, $C$ has the same value as that in
the first situation given by Eq. (5), while for $Q$, we have
\begin{eqnarray}
Q(\rho_{s_1s_2}(t))&=&-\text{H}(|\xi|^2) +
\text{H}(|\alpha\xi|^2)\nonumber\\
&+& \text{H}(\frac12 (1+\sqrt{1-4|\beta\xi\chi|^2})).
\end{eqnarray}
For reservoirs $\rho_{r_1r_2}$, we obtain
\begin{eqnarray}
C(\rho_{r_1r_2}(t))&=&\text{H}(|\beta\chi|^2)-\text{H}(\frac12(1-\sqrt{1-4|\beta\xi\chi|^2})),\nonumber\\
Q(\rho_{r_1r_2}(t))&=&\text{H}(\frac12(1+\sqrt{1-4|\beta\xi\chi|^2}))-\text{H}(|\chi|^2)\nonumber\\
&+&\text{H}(|\alpha\chi|^2).
\end{eqnarray}

With flat spectral density, we obtain the parameters in Eq. (8).
Concurrence of the spin system reads as $|\alpha^*\beta|e^{-\gamma
t}$, and therefore, there is no ESD. As time approaches infinity,
$Q$ can be represented asymptotically by
\begin{equation}
\lim_{t\rightarrow\infty}Q(\rho_{s_1s_2}(t)) \sim Q(0) e^{-\gamma
t},
\end{equation}
with $Q(0) = \text{H}(|\alpha|^2)$.

For this kind of initial state, $Q$ is no longer equal to $C$ as in
Fig. 3(a), but the correlations fall to zero monotonically and
continuously. For reservoirs, both correlations approach a definite
value monotonically  as in Fig. 3(b). The transference processes of
quantum and classical correlations are similar to those of the
initial state with two excitations, as shown in Figs. 3(c) and 3(d).
As time tends to infinity, both $Q$ and $C$ initially stored in the
spin system run into reservoirs. In addition, there are no increases
in the correlations.

For the Lorentz form in Eq. (11), we have the same parameters as in
Eq. (12). Concurrence of the spin system is given by
$|\alpha^*\beta\xi^2|$. $Q$ is not equal to $C$. As shown in Fig.
4(a), correlations of the spin system also oscillate. Figure 4(b)
shows evolutions of correlations in reservoirs. Figures 4(c) and
4(d) show evolutions for $s_1r_1$ and $s_1r_2$. Although there is
large oscillation due to the non-Markov effect, there are no
increases in correlations.

In conclusion, we have investigated the dynamics of both $Q$ and $C$
in the spin-boson model with two spins independently coupled to
their own reservoirs. From the analytical results obtained for $Q$
and $C$, we studied the evolutions of $Q$ and $C$ among different
partitions in detail. We found that the dynamics of $Q$ and $C$
depend closely on the form of the initial state. At the end of
evolution, all $Q$ and $C$ initially stored in the spin system
transfer to reservoirs. We found that $Q$ is more robust than
entanglement, for there is no sudden death with $Q$. At the same
time, during the evolution process, all partitions had nonzero $Q$,
which is not the case for entanglement. We also found that for a
large family of initial states, $Q$ remains equal to $C$ during the
course of evolution. In addition, there was no increase in either
correlation. There are many differences between this paper and
\cite{Ma09,W09}; for example, we focused on a more concrete case of
the spin-boson model, which is usually encountered, and we
concentrated on the transference processes of both $Q$ and $C$.

This work was supported by the National Fundamental Research
Program, National Natural Science Foundation of China (Grant Nos.
60121503 and 10874162).

\end{document}